\documentclass[nofootinbib,preprint,showpacs,preprintnumbers,amsmath,amssymb,aps,pre]{revtex4-2}
\usepackage[utf8]{inputenc} 
\usepackage{footnote}
\usepackage{epsfig}
\usepackage{natbib}
\usepackage{booktabs}
\usepackage{amsmath}
\usepackage{times}
\usepackage{epstopdf} 
\usepackage{graphicx}
\usepackage{color}

\usepackage{afterpage}

\begin{document}

\title{Stability enhanced by transient attractors in a memristive chaotic map}

\author{Alexandre R. Nieto}
\email[]{alexandre.rodriguez@urjc.es}
\affiliation{Nonlinear Dynamics, Chaos and Complex Systems Group, Departamento de
F\'{i}sica, Universidad Rey Juan Carlos, Tulip\'{a}n s/n, 28933 M\'{o}stoles, Madrid, Spain}

\author{Rubén Capeáns}
\affiliation{Nonlinear Dynamics, Chaos and Complex Systems Group, Departamento de
	F\'{i}sica, Universidad Rey Juan Carlos, Tulip\'{a}n s/n, 28933 M\'{o}stoles, Madrid, Spain}

\author{Miguel A.F. Sanju\'{a}n}
\affiliation{Nonlinear Dynamics, Chaos and Complex Systems Group, Departamento de
F\'{i}sica, Universidad Rey Juan Carlos, Tulip\'{a}n s/n, 28933 M\'{o}stoles, Madrid, Spain}

\date{\today}

\begin{abstract}

The introduction of a memristor in a chaotic system can significantly modify its dynamical behavior. In this paper, we couple a discrete memristor model with a chaotic map to investigate the memristor's impact on the stability of chaotic attractors. Our results reveal that introducing the memristor substantially enlarges the basin of attraction of a given chaotic attractor, thereby enhancing its stability. This phenomenon arises as a direct consequence of the memory and switching effects introduced by the memristor, which create new paths to the chaotic attractors.

\end{abstract}

\maketitle
\newpage

\textbf{The study of chaotic systems has been a central topic in nonlinear dynamics, with particular interest in understanding the stability and structure of chaotic attractors. Various approaches have been explored to analyze and manipulate chaos, including perturbations, control techniques, and modifications to system parameters. More recently, memristors, circuit elements with memory, have gained attention for their influence on dynamical behavior in both continuous and discrete systems. By incorporating a discrete memristor model into a chaotic map, we investigate how memory effects modify the stability of the system’s attractors. Our findings reveal that the memristor substantially enlarges the basin of attraction of chaotic attractors, demonstrating a fundamental mechanism by which memory and switching effects can play a constructive role in nonlinear dynamics.}

\section{Introduction} \label{sec1}
The memristor (resistor with memory) is the fourth fundamental electric component together with the capacitor, resistor, and inductor. A memristor is a resistor whose resistance depends on the electric current that has previously flowed through the device. It was first theorized by Leon Chua in $1971$ \cite{Chua71}, but it was not until $2008$ that a team at HP Labs successfully  achieved its first physical implementation \cite{Strukov08}. 

The mathematical model of the memristor is defined by
\begin{align}	
	\begin{aligned}
			&V(t)=M(q(t))I(t),  \\
		&\dot{q}=I(t),
		\label{Memristor}
	\end{aligned}
\end{align} 
where $V(t)$ is the voltage, $I(t)$ the current, $q(t)$ the charge, and $M(q(t))$ the charge-dependent resistance (i.e., the memristance). By applying the Euler method in the equation above, we derive a discrete memristor model of the form
\begin{align}	
	\begin{aligned}
		&V_{n}=M(q_n)I_n, \\
		&	q_{n+1}=q_n+hI_n,
		\label{Memristor2}
	\end{aligned}
\end{align} 
where $M(q_n)$ is a memristance function which usually includes an activation function such as the sigmoid. Henceforth, we set the time step to $h=1$ for simplicity.

In recent years, the effects of memristors in chaotic maps have garnered significant interest, particularly within the Chinese scientific community. The paper by Peng \textit{et al.} \cite{Peng20} is likely a pioneering work on this matter. Once the idea of coupling a discrete memristor model with a chaotic map was put on the table, numerous works studied the complex dynamics of this kind of system. References~\cite{Lei25,Li22,Rong22,Wang22} are some examples among many others. Here, we are not particularly interested in the complex dynamics that arise from introducing a memristor into a chaotic map, but rather in the qualitatively different dynamics that the memristor can generate.

For this research, we use the Lozi map as a model of a chaotic map and coupled it with the discrete memristor model described by Eq.~(\ref{Memristor2}). Our results, based on numerical simulations, show that the stability of the chaotic attractors is enhanced by the inclusion of the memristor. By enhanced stability, we mean that the basins of attraction of the chaotic attractors under the memristive Lozi map exhibit a significantly larger area. This phenomenon occurs because the memory and switching effects introduced by the memristor enable the orbits to find new paths to stability.

The structure of this paper is as follows. In Secs.~\ref{sec2} and \ref{sec3}, we introduce the memristive Lozi map, explain its expected properties, and define methods for a preliminary analysis of the system while avoiding the complexity introduced by the memristor. Numerical simulations of the memristive Lozi map are presented in Sec.~\ref{sec4}, where the enhancement of stability is demonstrated through the computation of basins of attraction. The underlying mechanisms responsible for this enhanced stability are explained in Sec.~\ref{sec5}. Finally, in Sec.~\ref{sec6}, we summarize the main results and discuss future perspectives on this topic.	

\section{A memristive Lozi map}\label{sec2}
The Lozi map is a well-known Hénon-like map introduced in 1978 by René Lozi \cite{Lozi78}. The properties of this map have been applied in a wide variety of fields, including chaos, synchronization, and electronic devices such as memristors (we refer the reader to Ref.~\cite{Lozi23} for a recent review on the map and its applications). The map is given by
\begin{align}	
	\begin{aligned}
		&x_{n+1}=1-a|x_n|+y_n, \\
		&y_{n+1}= bx_n,
		\label{Lozi}
	\end{aligned}
\end{align} 
where $a,b\in\mathbb{R}$ are constant parameters.

By coupling the Lozi map with the discrete memristor model given by Eq.~(\ref{Memristor}), we obtain a memristive Lozi map ($L_M$) of the form
\begin{align}
L_M:	\begin{cases}		
	x_{n+1}=1-a|x_n|+y_n, \\
	y_{n+1}= bx_n+M(z_n)y_n,\\
	z_{n+1}=y_n + z_n.
	\end{cases}		\label{EqM1}
\end{align} 

In these equations, $z$ represents the inner state of the memristor. The memristance function is $M(z_n)=k\tanh(z_n)$, where $k\in\mathbb{R}$ is the coupling strength between the discrete memristor model and the Lozi map. We have chosen $\tan(\cdot)$ as the activation function because it is widely used in the context of memristive systems and neural networks, owing to its smoothness, symmetry, and bounded range. However, any threshold function with similar properties could be used. We also emphasize that the way we have chosen to couple the maps is arbitrary, so other choices could be valid as well. The key idea is that we introduce a memory and a switch effect into the Lozi map.

In Eq.~(\ref{EqM1}), the variable $z$ accumulates the values of $y_n$ inside $\tanh(\cdot)$. Thus, we can reduce the dimensionality of the problem by replacing $z_n$ with a summation operator. With this change, Eq.~(\ref{EqM1}) can be rewritten as
\begin{align}
	L_M:\begin{cases}		
		x_{n+1}=1-a|x_n|+y_n, \\
		y_{n+1}= bx_n+My_n,\\
	\end{cases}		 	\label{EqM2}
\end{align} 
where now the memristance function is $M=k\tanh \left(z_0+\sum\limits_{j=0}^{n-1}y_j\right)$.  

As has been reported in many papers, including a memristor in discrete systems can generate complex dynamics such as chaos, multistability, bifurcations, hidden attractors, and more. The aim of this paper is not to comprehensively study the various dynamical behaviors that the system can exhibit for different parameter combinations. Therefore, we will consider $a=1$ because this simplifies the calculations, and $b=-0.05$ for convenience. We also set $z_0=0$, which means that no current flowed through the memristor at $n=0$. 

As we iterate Eq.~(\ref{EqM2}) from some initial condition $(x_0,y_0)$, the summation inside $\tanh(\cdot)$ will eventually diverge, causing $\tanh(\cdot)$ to converge to either $-1$ or $1$. Therefore, after a  large enough number of iterations, $M$ converges to the constant value $k$ or $-k$. 

For simplicity, while illustrating the following idea, we assume $k>0$ and that $L_M$ has a unique attractor $\mathcal{A}_+$ for $M=k$ and another unique attractor $\mathcal{A}_-$ for $M=-k$. The stability of the attractors depends on the average value of the $y-$coordinate within the attractor, $\bar{y}(\mathcal{A})$. If $\bar{y}(\mathcal{A}_+)>0$, $\mathcal{A}_+$ is stable because $M$ converges to $k$. On the other hand, if $\bar{y}(\mathcal{A}_+)<0$, $\mathcal{A}_+$ is unstable because $M$ converges to $-k$. In general, if $\text{sgn}[\bar{y}(\mathcal{A}_+)]\neq\text{sgn}[\bar{y}(\mathcal{A}_-)]$, either both or neither of the attractors is stable, leading to multistability or infinite oscillations between the attractors. On the other hand, if $\text{sgn}[\bar{y}(\mathcal{A}_+)]=\text{sgn}[\bar{y}(\mathcal{A}_-)]$, only one of the attractors is stable.

Regardless these analyses, the key idea for now is that we can anticipate the potential asymptotic states of $L_M$ by studying the attractors of an asymptotic map $A$. This map is obtained by setting $M=k$ in Eq.~(\ref{EqM2}), yielding
\begin{align}
A:	\begin{cases}		
		x_{n+1}=1-a|x_n|+y_n, \\
		y_{n+1}= bx_n+ky_n.
	\end{cases}		 	\label{EqA}
\end{align} 

Strictly speaking, the sign in front of $k$ in Eq.~(\ref{EqA}) could also be negative (i.e., if the case $M=-k$ occurs). However, this is not relevant since we will study the map for positive and negative values of $k$. 

\section{Analysis of the asymptotic map}\label{sec3}
\begin{figure}[h!]
	\centering
	\includegraphics[clip,height=11.5cm,trim=0cm 0cm 0cm 0cm]{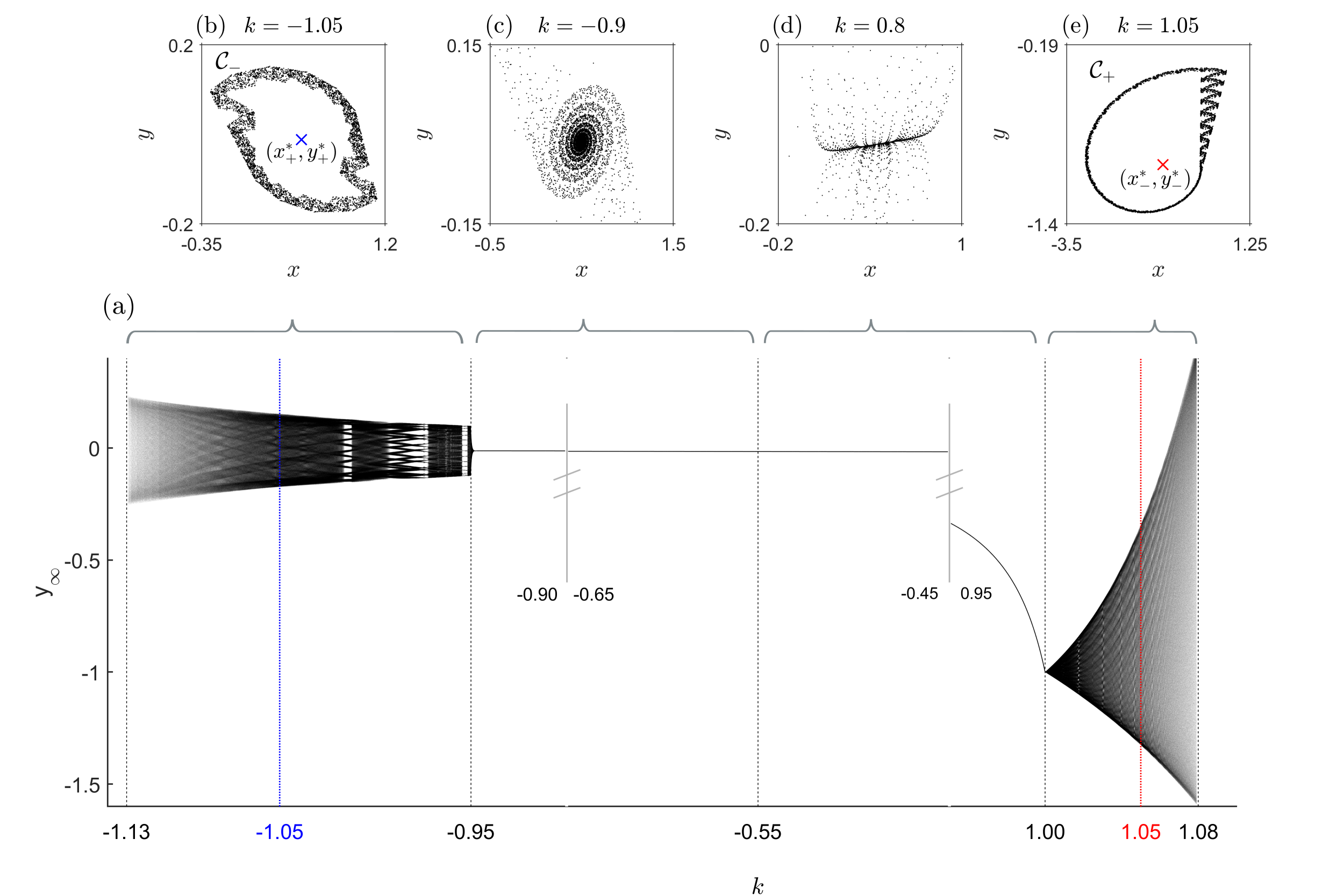}								
	\caption{(a) Bifurcation diagram of the asymptotic map given by Eq.~(\ref{EqA}), showing all the attractors existing for different values of $k$. Note that, in order to properly observe the bifurcations, there are two jumps in the value of $k$, so the fixed point is not represented in the intervals $k\in(-0.9,-0.65) \cup (-0.45,0.95)$. Panels (b-e) show particular examples of the qualitatively different attractors that the system can possess: (c) spiral, (d) node, and (b,f) chaotic attractors. The only stable fixed point of the system, $(x^*_+,y^*_+)$, loses stability to the left through a Neimark-Sacker bifurcation and to the right through a border-collision pair bifurcation. After each of these bifurcations, a chaotic attractor emerges. For $k<-0.95$, the attractor $\cal{C}_-$ surrounds $(x^*_+,y^*_+)$, while for $k>1$ the attractor $\cal{C}_+$ surrounds $(x^*_-,y^*_-)$. }
	\label{Fig1} 
\end{figure}

In this section, we analyze the dynamics of the asymptotic map, knowing that for a given value of $k$ the potential attractors of $L_M$ are the stable attractors of the asymptotic map with parameter $k$ or $-k$. All the possible attractors of the asymptotic map are represented in the bifurcation diagram shown in Fig.~\ref{Fig1}. For $k\in(-0.95,1)$ there exists a stable fixed point 
\begin{align}
	(x^*_+,y^*_+)=\left(\frac{1-k}{2.05-2k},cx^*_+\right),
\end{align} 
where $c=0.05/(k-1)$. This fixed point is a node for $k>-1+\sqrt{5}/5\approx-0.55$, and otherwise is a spiral. The system has a second fixed point 
\begin{align}
	(x^*_-,y^*_-)=\left(\frac{1-k}{0.05},cx^*_-\right),
\end{align} 
which only exists for $k>1$ and is unstable whenever it exists. 

To the left of $k=0$, $(x^*_+,y^*_+)$ loses stability through a Neimark-Sacker bifurcation (i.e., both eigenvalues of the Jacobian matrix at $(x^*_+,y^*_+)$ are complex and have unit modulus). At the bifurcation point, an attractive invariant curve is born. For lower values of $k$, the invariant curve undergoes a torus breakdown, becoming a chaotic attractor $\cal{C}_-$. This attractor, interrupted by periodic windows, exists for $k\in[-1.13,-0.95)$. For values of $k$ to the left of this interval, all initial conditions generate unbounded orbits. 

Another relevant bifurcation point is $k=1$. For this parameter value several things happen. First, the fixed point $(x^*_+,y^*_+)$ loses stability and becomes a saddle point. Second, $(x^*_-,y^*_-)$ appears for the first time, and it is an unstable node. Finally, a chaotic attractor $\cal{C}_+$ emerges surrounding $(x^*_-,y^*_-)$ after a border-collision pair bifurcation \cite{Zeraoulia11}. $\cal{C}_+$ is continuous in the interval $k\in(1,1.08]$, so is an example of robust chaos \cite{Banerjee98}. Again, to the right of this interval all the orbits are unbounded. 

In summary, the asymptotic map has a unique attractor for each value of 
$k$, implying that all initial conditions outside the basin of attraction are unbounded. Depending on the parameter value, the attractor may be a node, a spiral, an invariant curve, or a chaotic attractor ($\cal{C}_-$ or $\cal{C}_+$). 

From Fig.~\ref{Fig1}, we infer that the nature of the two potential attractors of $L_M$ can be arbitrarily chosen by defining an appropriate value of $k$. For example, for $|k|<-1+\sqrt{5}/5$ the potential attractors are nodes. For $|k|\in [-1+\sqrt{5}/5,0.95)$ one attractor is a node and the other is a spiral. A particularly interesting choice is $|k|\in(1,1.08]$, where both potential attractors are chaotic. Therefore, for the remainder of this paper, we will consider $k=1.05$ or $k=-1.05$, which fall within this interval.  

\begin{figure}[h!]
	\centering
		\includegraphics[clip,height=6.3cm,trim=0cm 0cm 0cm 0cm]{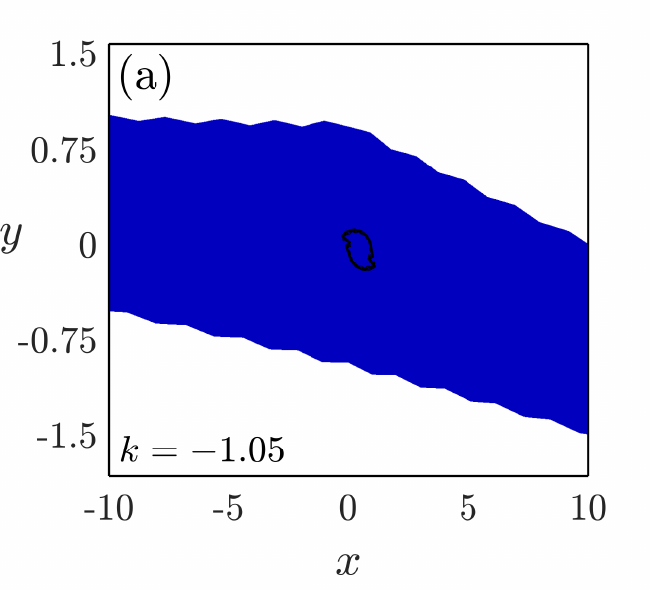}	
	\includegraphics[clip,height=6.3cm,trim=0cm 0cm 0cm 0cm]{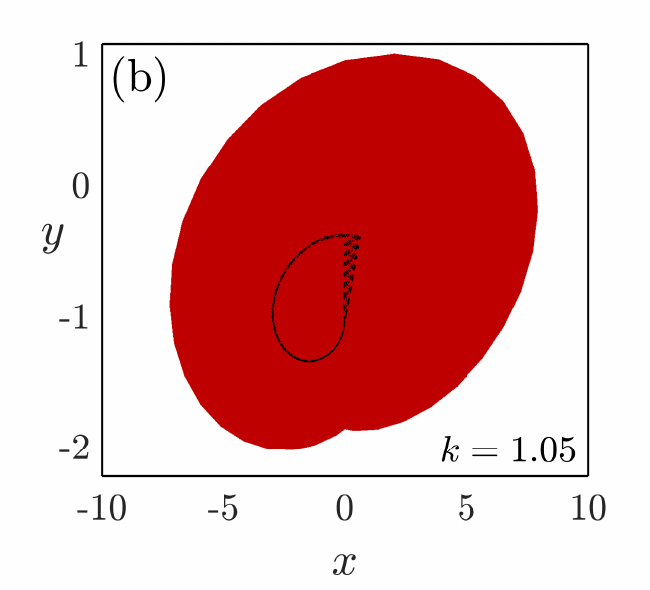}
	\caption{Basin of attraction of the chaotic attractors  (a) $\cal{C_-}$ and (b) $\cal{C_+}$ in the asymptotic map. Blue (red) color represents initial conditions within the basin of attraction of $\cal{C_-}$ ($\cal{C_+}$), while white represents initial conditions that lead to unbounded orbits. In each panel, the corresponding chaotic attractor is represented with black dots. }
	\label{Fig2} 
\end{figure}

We are interested in studying the stability of the chaotic attractors in the asymptotic map, so that we can later examine how their stability is modified by the memristor. We have quantified the stability of the attractors by computing the area of their basins of attraction in the phase space. The basins of attraction of the chaotic attractors $\cal{C_-}$ and $\cal{C_+}$ are shown in Fig.~\ref{Fig2}. To intuitively gauge the size of the basins, each panel is accompanied by the corresponding chaotic attractor. The area of the basins of attraction is $\mu[{\cal{B}}_A({\cal{C}}_-)]=188.6$ and $\mu[{\cal{B}}_A({\cal{C}}_+)]=34.9$, where the subscript $A$ indicates that the basin of attraction has been obtained iterating the asymptotic map.  

\section{Stability in the memristive map}\label{sec4}
In this section, we study the stability of the chaotic attractors in the memristive Lozi map defined by Eq.~(\ref{EqM2}). We have already seen that in the asymptotic map, $\cal{C_-}$ is stable for $k=-1.05$, while $\cal{C_+}$ is stable for $k=1.05$. However, for a given value of $k$, a stable attractor of $A$ is not necessarily stable in $L_M$. As we discussed in Sec.~\ref{sec2}, the stability of the attractors in the memristive map depends on the average value of their $y-$coordinate. In our case, $\bar{y}({\cal{C_-}})=-0.012$ and $\bar{y}({\cal{C_+}})=-0.849$. Since $\text{sgn}[\bar{y}({\cal{C_-}})]=\text{sgn}[\bar{y}({\cal{C_+}})]$, one and only one of the attractors is asymptotically stable for a given value of $k$. Specifically, $\cal{C_-}$ is the only attractor of $L_M$ when $k=1.05$, while $\cal{C_+}$ is the only attractor when  $k=-1.05$. This implies that, given our choice of parameters $a$ and $b$, $L_M$ does not exhibit multistability or infinite oscillation between attractors. We refer to attractors as ``asymptotically stable" rather than simply ``stable" because, as we will show later, an orbit may spend long transients within an attractor before eventually switching to the asymptotically stable attractor. 

\begin{figure}[h!]
	\centering
	\includegraphics[clip,height=8.5cm,trim=0cm 0cm 0cm 0cm]{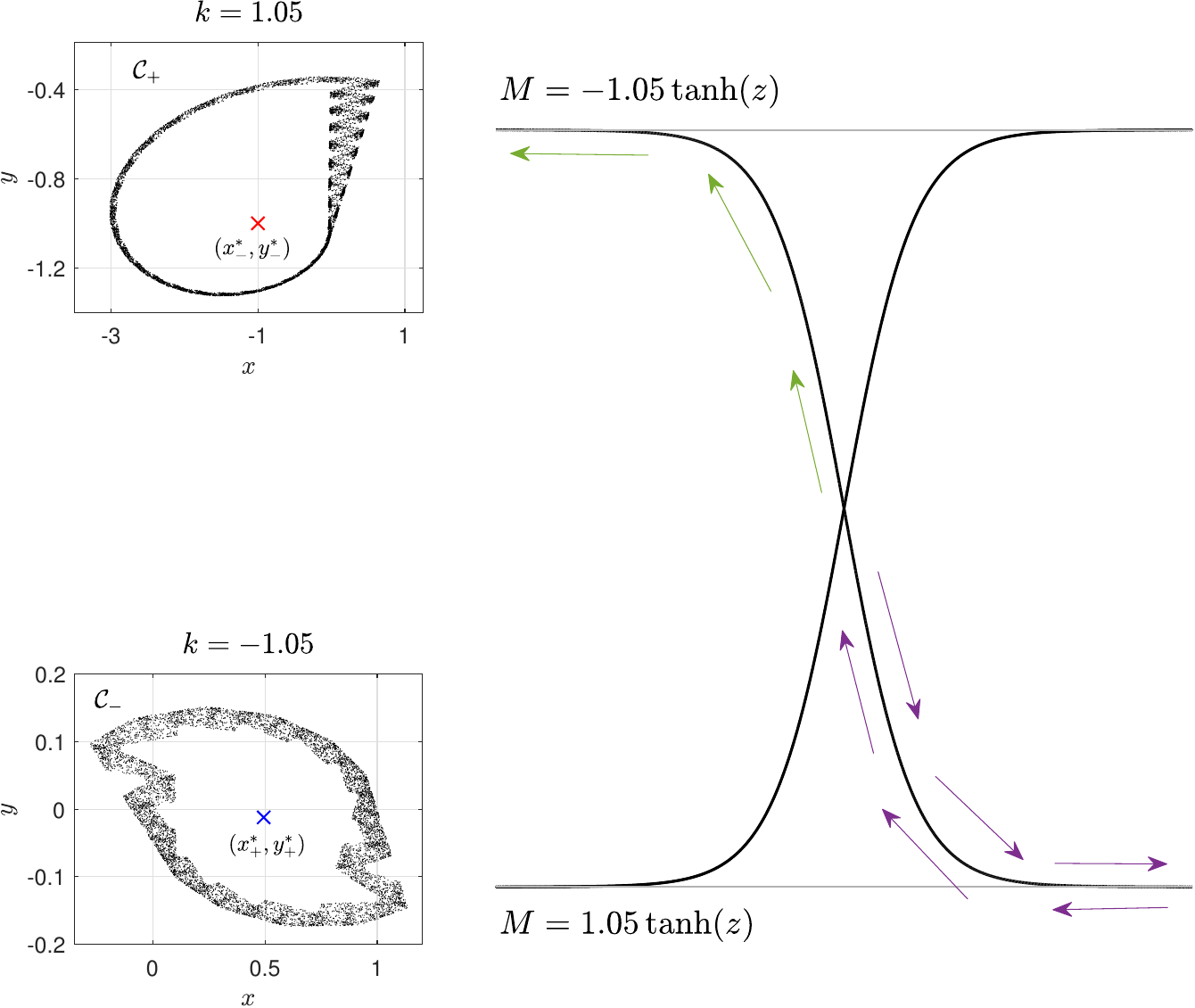}
	\caption{Basic dynamics in the memristive Lozi map given by Eq.~(\ref{EqM2}), for $|k|=1.05$. On the left side of the figure, the potential attractors are shown. On the right side, the memristance function for the positive and negative values of $k$ is displayed. The green arrows represent a case ($k=-1.05$) where $M$ directly converges to $1.05$, value for which $\cal{C}_+$ is asymptotically stable. The purple arrows represent an alternative path where the orbit visits the asymptotically unstable attractor $\cal{C}_-$ before ultimately converging to $\cal{C}_+$.}
	\label{Fig3} 
\end{figure}

The scheme shown in Fig.~\ref{Fig3} illustrates the basic dynamics of $L_M$ for $k=1.05$ and $k=-1.05$. On the left side of the figure, we display the potential attractors of $L_M$, which have been computed by iterating the asymptotic map. On the right side, we show the memristance function $M$ for the positive and negative values of $k$. For $k=-1.05$, all bounded orbits converge to $\cal{C}_+$, while for $k=1.05$ they converge to $\cal{C}_-$. One can easily imagine these paths by following  $M$ from $z=0$ to the corresponding attractor. However, it is important to note that there is no reason to assume that the value of $z$ either decreases or increases monotonically (see green arrows in Fig.~\ref{Fig3}). On the contrary, paths in which $\tanh(z)$ approaches $\pm1$, then switches sign, and subsequently converges to the opposite attractor are also possible. This particular path, for the case $k=-1.05$, is represented in Fig.~\ref{Fig3} by purple arrows followed by green ones. We will discuss this scenario in more detail in the following section.

\begin{figure}[h!]
	\centering
	\includegraphics[clip,height=6.3cm,trim=0cm 0cm 0cm 0cm]{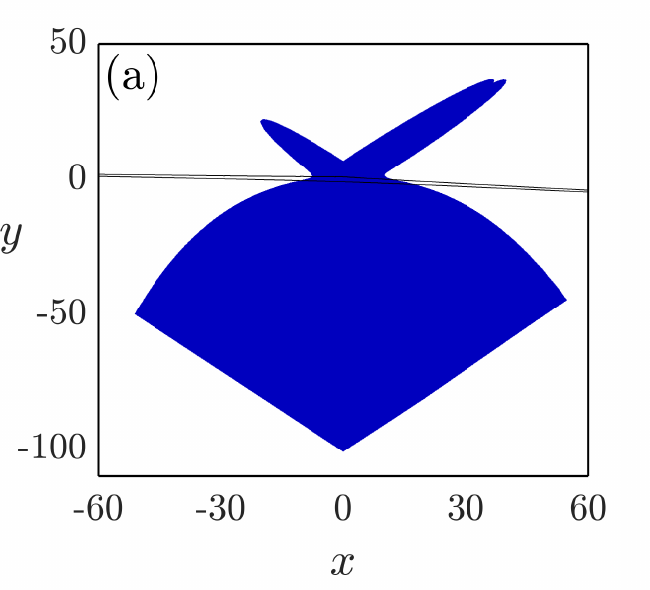}
	\includegraphics[clip,height=6.3cm,trim=0cm 0cm 0cm 0cm]{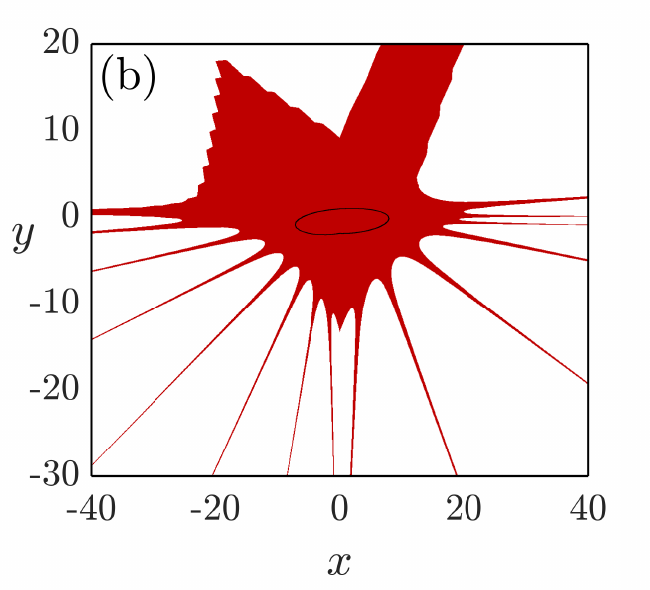}			
	\caption{ Basin of attraction of (a) $\cal{C_-}$ for $k=1.05$ and (b) $\cal{C_+}$ for $k-=1.05$ in the memristive Lozi map given by Eq.~(\ref{EqM2}). As in Fig.~\ref{Fig2}, blue (red) color represents initial conditions within the basin of attraction of $\cal{C_-}$ ($\cal{C_+}$). White represents initial conditions that generate unbounded orbits. In each panel, the boundary of the basin of the chaotic attractor in the asymptotic map is represented using dark lines.  }
	\label{Fig4} 
\end{figure}

Once we have established an understanding of the possible behaviors of an orbit under $L_M$, we proceed to analyze the stability of the chaotic attractors through numerical simulations. For this purpose, we have computed the basins of attraction of $\cal{C}_+$ and $\cal{C}_-$ under $L_M$, using $k=1.05$ and $k=-1.05$. The results are shown in Fig.~\ref{Fig4}, where the color scheme is as in Fig.~\ref{Fig2}. For comparison, the boundaries of the basins in the asymptotic map are overlaid on the basins of the memristive map. The size difference is striking. The area of the basins of attraction is $\mu[{\cal{B}}_{LM}({\cal{C_-}})]=7114.0$ and $\mu[{\cal{B}}_{LM}({\cal{C_+}})]=1787.8$. Compared to the asymptotic map, the basin of $\cal{C_-}$ is $38$ times larger, while that of $\cal{C_+}$ is $51$ times larger. These results demonstrate that the stability of the attractors is significantly enhanced by introducing the memristor. In the next section, we will explain the mechanisms responsible for this enhancement.

\section{Routes to stability}\label{sec5}
The presence of the memristor makes it challenging to develop intuitive insights about the behavior of the memristive Lozi map $L_M$. As a result, the significant enlargement of the basins might seem counterintuitive. To explain this phenomenon, we consider an initial condition that generates an unbounded orbit in the asymptotic map. When the same initial condition is chosen in the memristive map, the orbit initially lies outside the basin of attraction of the chaotic attractor. However, during the first iterations (before $\tanh(z)$ reaches extreme values), $|M|$ is low, causing the orbit to move towards the fixed point. Since the basins of attraction of these fixed points are large, many initial conditions are attracted to them. 

As $\tanh(z)$ approaches extreme values $\pm 1$, the fixed points that previously attracted the orbits become unstable. Nevertheless, their positions are generally within the basin of attraction of $\cal{C_-}$ or $\cal{C_+}$. This is not a coincidence, but rather a direct consequence of the position of the chaotic attractors surrounding the fixed points. For $\cal{C_-}$, as $M$ decreases, the fixed point $(x_+^*,y_+^*)$ smoothly guides the orbit towards the attractor. The situation for $\cal{C_+}$ is slightly different. As $M$ increases, $(x_+^*,y_+^*)$ guides the orbit until $M=1$. At this point, $(x_+^*,y_+^*)$ becomes unstable, but in the same bifurcation $\cal{C_+}$ appears and attracts the orbit. Furthermore, for parameter values where $\cal{C_+}$ exists, $(x_+^*,y_+^*)$ is a saddle whose unstable manifold naturally expels the orbits directly towards $\cal{C_+}$. 

This scenario might seem like a mere pleasant coincidence, but the truth is that most of the characteristics we have described are typical of chaotic systems. Chaotic attractors rarely emerge spontaneously, but they appear through bifurcations involving the formation and destruction of fixed points. In fact, bifurcation diagrams like the one shown in Fig.~\ref{Fig1} are among the most common patterns in chaos theory.

\begin{figure}[h!]
	\centering
	\includegraphics[clip,height=7.2cm,trim=0cm 0cm 0cm 0cm]{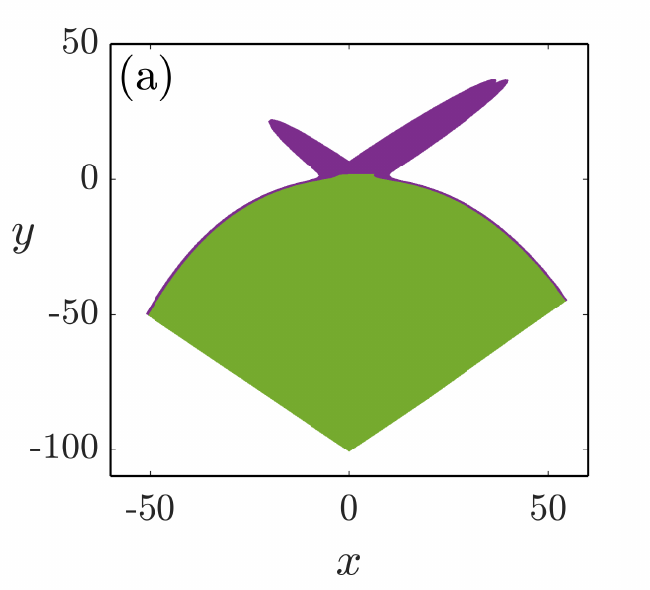}
	\includegraphics[clip,height=7.2cm,trim=0cm 0cm 0cm 0cm]{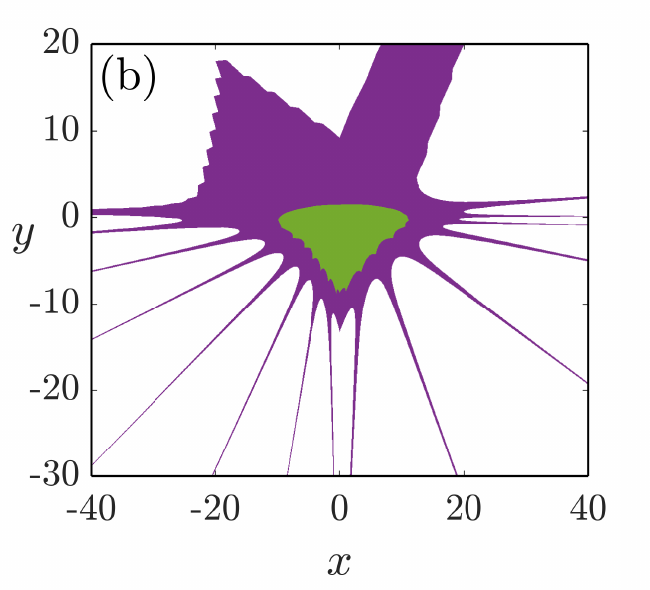}
	\caption{Same basins of attraction depicted in Fig.~\ref{Fig4}, but now classifying the initial conditions within the basins depending on the route to the attractor followed by the orbits: $(x^*_+,y^*_+)\to{\cal{C}}_s$ (green) or $(x^*_+,y^*_+)\to{\cal{C}}_u\to(x^*_+,y^*_+)\to{\cal{C}}_s$ (purple). The	subscript of $\cal{C}$ stands for asymptotically unstable ($u$) or asymptotically stable ($s$). }
	\label{Fig5} 
\end{figure}

In the memristive Lozi map, not all orbits follow the same path to the chaotic attractor, but orbits can take multiple routes to stability. These orbits can be broadly classified into two fundamental types: those that directly approach the stable attractor via the fixed points, and those that initially spend some time in the asymptotically unstable chaotic attractor before reaching the stable attractor. If we observe again Fig.~\ref{Fig3}, we see that the first route to stability is represented by green arrows, while the second route starts with the purple arrows and continues with the green ones. Using the same color scheme, Fig.~\ref{Fig5} shows the basins of attraction of Fig.~\ref{Fig4} again, but this time classifying the initial conditions based on whether their orbits pass through the asymptotically unstable attractor or not.

\begin{figure}[h!]
	\centering
	\includegraphics[clip,height=11cm,trim=0cm 0cm 0cm 0cm]{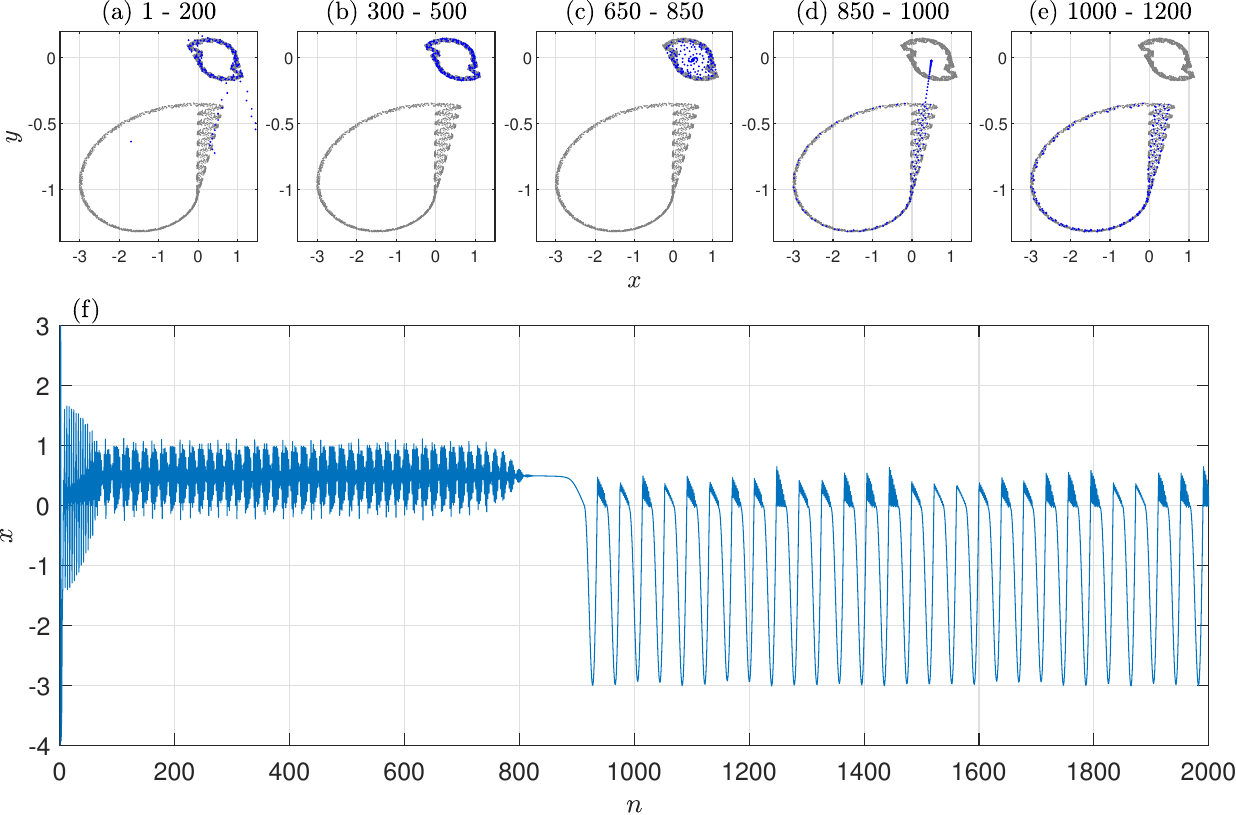}
	\caption{An exemplary orbit in the memristive Lozi map for $k=-1.05$. Panels (a-e) represent different stages of the orbit's evolution: (a) initial approach to the asymptotically unstable attractor $\cal{C}_-$, (b) long transient within $\cal{C}_-$, (c) transition from ${\cal{C}}_-$ to the fixed point $(x^*_+,y^*_+)$, (d) transition from $(x^*_+,y^*_+)$ to the stable attractor ${\cal{C}}_+$, and (e) asymptotic behavior in $\cal{C}_+$. Panel (f) presents the entire evolution of the orbit, culminating in its stabilization in $\cal{C}_+$. The low-amplitude chaotic oscillations for $n\lesssim800$ correspond to the transient dynamics within $\cal{C}_-$, while the high-amplitude oscillations for $n\gtrsim 900$ correspond to the the orbit's stabilization in $\cal{C}_+$.   }
	\label{Fig6} 
\end{figure}

We will conclude our results by delving deeper into a particularly surprising case: an orbit that spends a long transient within an attractor that is asymptotically unstable. For this numerical experiment, we set $k=-1.05$ and select an arbitrary initial condition from the purple region of Fig.~\ref{Fig5}(b). The complete evolution of the $x-$coordinate of the orbit is depicted in Fig.~\ref{Fig6}(f). In the upper part of the figure [see panels (a-e)], we illustrate different stages of the orbit's evolution. In each panel, the stable attractors of the asymptotic map are plotted with gray dots, while some points of the orbit are represented in blue. It is important to note that these attractors do not coexist for any given parameter value. They are represented together solely to provide a clearer understanding of the dynamics.

After the first iterations, a large positive value (e.g., $10$) accumulates inside $\tanh(\cdot)$. As a consequence, $M=-1.05$ and the orbit quickly approaches $\cal{C}_-$ [see Fig.~\ref{Fig6}(a)], which is a stable attractor of $A$ for $k=-1.05$. As the iteration of $L_M$ progresses, the summation inside $\tanh(\cdot)$ decreases slowly (recall that $\bar{y}({\cal{C}_-})<0$), causing the orbit to spend a long transient within $\cal{C}_-$ [see panel (b)]. Once the positive reservoir inside $\tanh(\cdot)$ vanishes, $\cal{C}_-$ loses stability, and the orbit transitions to the fixed point $(x^*_+,y^*_+)$, which is a spiral [see panel (c)]. As $M$ continues increasing, $(x^*_+,y^*_+)$ becomes unstable and expels the orbit towards $\cal{C}_+$ [see panel (d)]. Since $\bar{y}({\cal{C}_+})<0$, $\lim\limits_{n\to\infty}z_n=-\infty$ and $\lim\limits_{n\to\infty}M=1.05$, making the chaotic attractor asymptotically stable. Thus, the situation plotted in Fig.~\ref{Fig6}(e) represents the orbit's asymptotic behavior.

\section{Conclusions and discussion}\label{sec6}

In this paper, we have shown that the introduction of a memristor can enhance the stability of a chaotic attractor by significantly increasing the area of its basin of attraction. We have illustrated the main ideas using the memristive Lozi map as a model, but we believe the underlying phenomena are quite general. Thus, similar results could likely be observed in other continuous or discrete systems. While the way we introduced the memristor in the model is arbitrary, its effects on the dynamics are not a consequence of this particular choice. 

For the enhancement of stability to occur, the dynamical system must have three main properties. First, the basin of attraction of the chaotic attractor (or the attractor whose stability we want to enhance) must have a relatively tiny basin of attraction. Otherwise, the idea has no purpose. Second, the system must possess stable attractors with large basins of attraction for parameter values close to $0$. These attractors will act as transient attractors that initially retain the orbits. Finally, as the parameter varies, paths connecting the transient attractors to the stable attractor must exist. In many dynamical systems, the existence of these paths is a direct consequence of the bifurcations. 

We have focused our work on this particular phenomenon, but during this research we have found different mechanisms that might also be of interest for future research. In particular, the presence of infinite oscillations between transient attractors and the existence of multistability can arise in systems where $\text{sign}({\cal{A_+}})\neq\text{sign}({\cal{A_-}})$. Also, here we have considered that the memory of the memristor is infinite (i.e., the number of values of $y$ accumulated inside the memristance function is not bounded). However, by choosing finite memory effects new dynamical behaviors might arise. Finally, the choice of $\tanh(\cdot)$ as an activation function is common and serves its purpose. However, the slope of the memristance function around zero is what determines the velocity of the transition between the two possible asymptotic states. By choosing different functions with different slopes, a higher or lower enhance of stability might be achieved. 

 \section*{ACKNOWLEDGMENTS}
This work has been financially supported by MCIN/AEI/10.13039/501100011033 and by “ERDF A way of making Europe” (Grant No. PID2023-148160NB-I00).

\section*{AUTHOR DECLARATIONS}

\textbf{Conflict of Interest}

The authors have no conflicts to disclose.\\

\textbf{Author Contributions}

\textbf{Alexandre R. Nieto:} Conceptualization (equal); Formal Analysis (lead); Methodology (lead); Software (lead); Validation (supporting); Visualization (lead); Writing – original draft (lead); Writing – review and editing (equal).
\textbf{Rubén Capeáns:} Conceptualization (equal), Formal Analysis (supporting); Methodology (supporting); Software (supporting); Validation (lead); Visualization (lead); Writing – review and editing (equal).
\textbf{Miguel A.F. Sanjuán:} Conceptualization (equal); Funding Acquisition (lead); Methodology (supporting); Writing – review and editing (equal).

\section*{DATA AVAILABILITY}

The data that support the findings are available from the corresponding author upon reasonable request.

\end{document}